\documentclass[a4paper,10pt]{article}
\usepackage[utf8]{inputenc}

\title{A   Holographic Bound  for   D3-Brane}
\author{ Davood Momeni$^{1}$\footnote{Corresponding author }, Mir Faizal$^{2, 3}$,Aizhan Myrzakul$^{4}$ , \\Sebastian Bahamonde$^{5}$ , Ratbay Myrzakulov$^{4}$ 
\\\\$^1$  Department of Physics, College of Science, Sultan Qaboos University , \\P.O. Box 36, Alkhod 123, Oman
 \\$^2$ Irving K. Barber School of Arts and Sciences, 
\\ University of British
Columbia - Okanagan,  
3333 University Way,\\  Kelowna,   British Columbia V1V 1V7, Canada
\\$^3$ Department of Physics and Astronomy, University of Lethbridge,\\
Lethbridge, Alberta, T1K 3M4, Canada 
\\$^4$
Eurasian International Center for Theoretical Physics \\
and Department of General Theoretical Physics, \\
Eurasian National University, Astana 010008, Kazakhstan
\\ $^5$
Department of Mathematics, University College London,\\  Gower Street, London, WC1E 6BT, UK
}
\date{}  
\begin{document}

\maketitle

\begin{abstract}

In this paper, we   will regularize the holographic entanglement entropy, 
holographic complexity and fidelity susceptibility for a configuration of  D3-branes.
We will also study the regularization of the holographic complexity from action for 
a configuration of  D3-branes.
It will be demonstrated 
that for a spherical shell of D3-branes
the regularized holographic  complexity  is always greater than or equal to than the
regularized fidelity susceptibility. Furthermore, we will 
also demonstrate that the regularized holographic complexity is related to the regularized 
holographic entanglement entropy for this system. 
Thus, we will obtain a holographic bound involving   regularized 
holographic complexity, regularized holographic entanglement entropy and regularized 
fidelity susceptibility 
of a configuration of D3-brane.  We will also discuss a bound for regularized  holographic 
complexity from action, for a D3-brane configuration. 
\end{abstract}

In this paper, we will analyse the relation between the holographic complexity,
holographic entanglement entropy and 
fidelity susceptibility for a spherical shell of D3-branes. 
We shall also analyze the holographic complexity form action for  a configuration 
of D3-branes. 
These quantities will be geometrically calculated using the bulk geometry, and  
the results thus obtained will be used to demonstrate the existence of a 
holographic bound for configurations of  D3-branes. 
It may be noted that there is a  close relation between the geometric configuration involving  
 D3-branes and  quantum informational systems  \cite{branes}. 
It is known that D3-branes can be analysed as  a real three-qubit state \cite{d3}.
This is done using the 
the configurations of intersecting D3-branes,  wrapping around the six compact dimensions.
The $T^6$
 provides the microscopic string-theoretic interpretation of the charges. The 
  most general real three-qubit state can be parameterized by four real 
  numbers and an angle, and that the most
general STU black hole can be described by four D3-branes intersecting at an angle.
Thus, it is possible to represent 
a three-qubit state by D3-branes. A  system  D3-branes have been used to
holographically analyse quantum Hall effect, 
as a system of D3-D7-branes has been used to obtain  the Hall 
conductivity and the topological entanglement entropy 
for quantum Hall effect \cite{qh}. The mutual
information between two  spherical regions
in $\mathcal{N}= 4$ super-Yang-Mills theory  dual to
type IIB string theory on AdS$_5 \times S^5$
has been analysed using correlators of surface operators \cite{ii}. Such a  surface operator  
 corresponds to having a D3-brane in AdS$_5 \times S^5$ ending on the boundary along 
 the prescribed surface. 
This construction relies on the strong analogies between the twist field operators  
used for the computation of the entanglement entropy, 
and the disorder-like surface operators in gauge theories. 
A a configuration of D3-branes and D7-branes with a non-trivial 
worldvolume gauge field on the D7-branes has also been used 
to holographically analyse new form of quantum liquid, with certain properties 
resembling a Fermi liquid \cite{ferm}
 The holographic entanglement entropy of an infinite strip subsystem on the 
 asymptotic AdS boundary
 has been  used as a probe to study the thermodynamic instabilities of 
 planar R-charged black holes 
 and  their dual field theories \cite{ther}.  This was done using a  
 spinning D3-branes with one non-vanishing angular momentum.
It was demonstrated that the  holographic entanglement entropy  
exhibits the thermodynamic instability associated 
with the divergence of the specific heat. When the width of the 
strip was large enough, the finite part of the holographic entanglement entropy as
a function of the temperature resembles the thermal entropy. 
However as  the width became smaller,   the two entropies behave differently. 
It was also observed that below   a critical value for the width of the strip, 
the finite part of the holographic entanglement entropy 
as a function of the temperature develops a self-intersection.  
 
Thus, there is a well established relation between different 
D3-branes configurations and information
theoretical processes. Thus, it would be interesting to analyze different 
information theoretical quantities for a configuration of D3-branes. 
It may be noted that entropy is one of the most 
important quantities in information theoretical processes. This is
because entropy measures the lose of information 
during a process. It may be noted  
that maximum entropy of a region of space scales with its area, 
and this observation has been motivated 
 from the physics of black holes.  This observation has   led to the
 development of the   holographic principle  \cite{1, 2}. 
 The holographic principle equates the   degrees of freedom in a 
 region of space   to the   degrees 
of freedom on the boundary surrounding that region of space.
The AdS/CFT correspondence is  a concrete 
realizations of the holographic principle \cite{M:1997}, and 
it relates the string theory  in  AdS 
to a  superconformal field theory on the boundary of that AdS. 
The AdS/CFT correspondence in turn can be used 
to  holographically obtain the  entanglement entropy of a boundary field theory. 
The holographic entanglement entropy 
 of a conformal field theory on the boundary of an AdS solution  
 is dual to the area of a minimal surface defined in the bulk. Thus, for a  
   subsystem as $A$, we can define    $\gamma_{A}$ as   the $(d-1)$-minimal surface extended 
into the AdS bulk,  with the boundary $\partial A$. Now using this subsystem,
the  holographic entanglement entropy can be expressed as \cite{6, 6a}
\begin{equation} \label{HEE}
S_{A}=\frac{\mathcal{A}(\gamma _{A})}{4G_{d+1}}
\end{equation}
where $G$  is the gravitational constant
for the bulk AdS and $\mathcal{A}(\gamma _{A})$ is the area of the minimal surface. 
Even though this  quantity is divergence, it can be regularized  \cite{u1, u2}. 
The holographic entanglement entropy can be regularized by subtracting the contribution
of the background AdS spacetime from the deformation of the AdS spacetime. Thus, for 
the system studied in this paper, let $ \mathcal{A} [D3 (\gamma _{A})]$ be 
the contribution of a  D3-brane shell  and $ \mathcal{A} [AdS (\gamma _{A})]$
 be contribution of the background AdS spacetime, then the regularized holographic entanglement 
 entropy will be given by 
\begin{equation}  
\Delta S_{A}=\frac{ \mathcal{A} [D3 (\gamma _{A})] - \mathcal{A} [AdS (\gamma _{A})]  }{4G_{d+1}}. 
\end{equation}
In this paper, we  will use this regularized holographic entanglement entropy. 

The entropy measures the loss of information during a process. However, it is also important to 
know how easy is it for observer to extract this information.  The complexity
 quantified this idea relating to  difficulty  to extract  information.  It is
 expected that complexity is another 
fundamental physical quantify, as it is an important quantity in information  theory,
and law of physics can be 
  represented in terms of informational theoretical processes. In fact, complexity has  
  been used in condensed matter systems \cite{c1, c2} and  molecular physics \cite{comp1, comp2}.   Complexity is also important in the    black hole physics, as it has been proposed 
  that even thought the   information may not be ideally lost during the evaporation of  a 
  black hole,   it  would be effectively 
lost during the evaporation of  a black hole. This is because  it would become  impossible
to reconstruct it from the Hawking radiation \cite{hawk}. 
It has been  proposed that   the 
complexity can be obtained   holographically as a quantity  
  dual to   a volume of  codimension one time 
slice in anti-de Sitter (AdS)
\cite{Susskind:2014rva1,Susskind:2014rva2,Stanford:2014jda,Momeni:2016ekm}, 
\begin{equation}\label{HC}
Complexity = \frac{V }{8\pi R G_{d+1}},
\end{equation}
where $R$ and $V$ are the radius of the curvature and the volume in the AdS bulk. 

As it is possible to define the volume in different ways in the AdS,
different proposals for the complexity have been made. 
If this volume is defined to be the maximum volume in AdS which
ends on the time slice at the AdS boundary,   $V = V(\Sigma_{max})$,
then the complexity corresponded to  fidelity susceptibility $\chi_F$
of the boundary conformal  field theory \cite{r5}. 
This quantity   diverges \cite{u4}. 
However, we will  regularize it by subtracting the contribution 
of the background AdS spacetime from the contribution of the deformation of AdS spacetime. 
So,    let $  V [D3 (\Sigma_{max})]$ be 
the contribution of a  D3-brane shell  and $ V [AdS (\Sigma_{max})]$
 be contribution of the background AdS spacetime, then we can write 
 the regularized  fidelity susceptibility as
\begin{equation}  
\Delta \chi_F=\frac{ V [D3 (\Sigma_{max})] 
- V [AdS (\Sigma_{max})]  }{4G_{d+1}}. 
\end{equation}
It is also possible use a subsystem $A$ (with its complement), to  define a  volume 
in AdS as $V = V(\gamma_A)$. This is the     volume which is 
  enclosed by the   minimal surface used to calculate the holographic 
entanglement entropy \cite{Alishahiha:2015rta}. Thus, using $V = V(\gamma_A)$, we obtain  the 
 holographic  complexity as 
$ \mathcal{C}_A$.  As we want to differentiate  between these two cases, we shall  call 
this the  quantity define by  $V = V(\Sigma_{max})$ as 
fidelity susceptibility, and  the quantity denied by $V = V(\gamma_A)$ as holographic complexity. 
The holographic complexity    diverges \cite{u4}. We will   regularized it
by subtracting the contributions of the background AdS from the deformation of the AdS spacetime. 
Now if  $  V [D3 (\gamma_A)]$ is 
the contribution of a  D3-brane shell  and $ V [AdS (\gamma_A)]$ is the contribution 
of the background AdS spacetime, then we can write the regularized holographic complexity as 
\begin{equation}  
\Delta  C_A =\frac{ V [D3  (\gamma_A)] - V [AdS  (\gamma_A)]   }{4G_{d+1}}. 
\end{equation} 

It may be noted that there is a different proposal for calculating the holographic complexity 
of a system using the action \cite{prl, prd}. According to this proposal 
 the holographic complexity of a system can be related to the   bulk action evaluated on
the Wheeler-deWitt patch, 
\begin{eqnarray}
\mathcal{C}_W=\frac{A(W)}{\pi \hbar},  
\end{eqnarray}
where $A(W)$ is the action evaluated on the Wheeler-DeWitt patch  $W$, 
with a suitable boundary time. 
 To differentiate it from the holographic complexity calculated from volume $\mathcal{C}$, 
 we shall call this 
 quantity ''holographic complexity from action'', and denote it by $\mathcal{C}_W$ 
 (as it has been calculated on a Wheeler-DeWitt patch). This quantity also diverges \cite{u5}.  
 We shall regularize it by subtracting the contributions 
 of  AdS spacetime from the contributions of the deformation of AdS spacetime. So, 
 if  $ A  [D3(W)] $ is 
the contribution of a  D3-brane shell  and $A [AdS(W)]  $ is the contribution 
of the background AdS spacetime, then we can write the regularized holographic complexity  from 
action as 
\begin{eqnarray}
\Delta \mathcal{C}_W=\frac{A[D3(W)]- A[AdS(W)]}{\pi \hbar}. 
\end{eqnarray}
It may be noted that this proposal is very different from the other proposals to calculate 
complexity of a boundary theory. This difference occurs as there are differences in the definition
of  complexity for a boundary field theory. So, this proposal cannot be directly related to 
the proposals where the complexity can be calculated from the volume of a geometry. In fact, 
it is possible to have the same volume for two theories with different field content. 
In this paper, we will first use calculate a bound for the D3-brane geometries using the 
volume of a shell of D3-branes. Then we shall calculate a different holographic bound 
for a configuration of D3-branes using the action of this system. 

In this paper, we will analyze a specific  configuration of D3-branes, and 
discuss  the behavior of these regularized information theoretical quantities for it. 
It is possible to use static gauge, and
write the bosonic part of the  action for such a system   
in  $AdS_5 \times S^5$
background  as  \cite{Schwarz:2014rxa}
\begin{equation}\label{Naction}
A = \frac{1}{2 \pi g_s k^2} \int  \left(\sqrt{-h}
- \sqrt{- \det \left( G_{\mu\nu} +  k F_{\mu\nu}\right)} \right)\, d^4 x
+ \frac{\chi}{8\pi} \int F\wedge F,\label{Action}
\end{equation}
where $k = \sqrt{g_s N/\pi}$ and 
\begin{equation}
G_{\mu\nu} = h_{\mu\nu} + k^2\frac{\partial_\mu \phi^I \partial_\nu \phi^I}{ \phi^2}.
\end{equation}
 Here  $h_{\mu\nu} = \phi^2 \eta_{\mu\nu}$,
$ h = \det h_{\mu\nu}$, with $\eta_{\mu\nu}$ being  the four dimensional Minkowski metric. 
Thus, we can write  $\sqrt{-h} = \phi^4$, where $\phi^2 = \sum(\phi^I)^2 $, and $\phi^I$ are 
 six scalar fields  corresponding  to the six dimensions transverse
to the D3-brane geometry. It may be noted that  $\int F \wedge F$ term only
contributes to the  magnetically charged configurations. 
The   D3-brane 
can be placed at a  fixed position on $S^5$, such that the  
five scalars fields  corresponding  to the $S^5$
  geometry will not have any contribution. We shall consider the 
 spherically symmetrical static solutions, centered at $r=0$,
 for this geometry. So, the 
  electric field $\vec{E}$ and the  magnetic fields   $\vec{B}$ will only
have  radial components, which we shall denote by $E$ and $B$. 
So, all  fields of this system 
are only functions of the radial coordinate $r$, $E(r)$, $B(r)$, $\phi(r)$. 
Thus, we can write 
$\det(-G_{\mu\nu}) = \phi^6 G_{rr}= \phi^6[\phi^2 + \gamma^2 (\phi'/\phi)^2],
$
and 
\begin{equation}
-\det (G_{\mu\nu} + \gamma F_{\mu\nu})  = \phi^6 \left(G_{rr} 
- \frac{\gamma^2 E^2}{\phi^2}\right)
\left( 1 + \frac{\gamma^2 B^2}{\phi^4} \right).
\end{equation}
So,    the Lagrangian density for this system can be written as 
\begin{equation}\label{gamma}
{\cal L} = \frac{1}{\gamma^2}  \phi^4\left(1 -
\sqrt{\left( 1 + \frac{\gamma^2 [(\phi')^2-E^2]}{\phi^4} \right)
\left(1 +  \frac{\gamma^2 B^2}{\phi^4}\right)} \, \right) +g_s \chi BE.
\end{equation}
where $\gamma=\sqrt{\frac{N}{2\pi^2}}=R^2\sqrt{T_{D3}}$, $T_{D3}$ is  D3-brane tension.
There  are two BPS solutions for this geometry, 
$
\phi_{\pm} = \mu \pm Q/r.
$ The probe D3-brane solution discussed here   describes a BIon like  spike  
(either up to the AdS$_5$ boundary or down to the Poincare horizon, depending on the sign 
in $\phi_{\pm}$). This  solution also breaks the 
 translational symmetry in the field theory, and preserves the rotational invariance. 
 
 It is also possible to analyze 
  a probe D3-brane with $Q=0, E=0, $
and $B=0$.  
Now we will analyze such a  specific solution representing a D3-brane configuration, 
and analyze these quantities for that specific geometric configuration. 
It is possible to study such a   D3-brane shell. The 
   metric for the near horizon geometry of D3-brane shell is given by \cite{Kraus:1998hv}
	\begin{eqnarray}
		&&ds^2=\frac{R^2}{z^2h(z)}\Big(\sum_{\mu=0}^{3}dx_\mu dx^{\mu}\Big)+R^2h(z)\Big(\frac{dz^2}{z^2}+d\Omega_5\Big)\,\label{metric}
	\end{eqnarray}
where the function $h(z)$ is defined as
	\begin{equation}
		h(z)= \left\{ \begin{array}{lr}
			1 \ , & z\leq z_0 \\
			(\frac{z_0}{z})^2\ , &z\geq z_0
		\end{array} \right. \ .
	\end{equation}
For this geometry, the entangled region is a strip with width $\ell$ in the D3-brane shell 
defined by the embedding $A=\{x=x(z),t=0\}$. The area functional can be expressed as
	\begin{eqnarray}
		&&\mathcal{A}(\gamma _{A})=2\pi^3R^8L^2\int_{0}^{z_{*}}\frac{h(z)\sqrt{x'(z)^2+h(z)^2}}{z^3}dz\,,
	\end{eqnarray}
where $x'(z_*)=\infty$. The Euler-Lagrange equation for $x(z)$ has the following form
	\begin{eqnarray}
		&&\frac{x'(z)}{\sqrt{x'(z)^2+h(z)^2}}=\frac{h(z_*)}{h(z)}\Big(\frac{z}{z_*}\Big)^3\label{xD3}
	\end{eqnarray}
The total length   can be obtained by 
	\begin{eqnarray}
		\ell &=&2\int_{0}^{z_*}dzh(z)
		\displaystyle\Big[\frac{\frac{h(z_*)}{h(z)}\big(\frac{z}{z_*}\big)^3}{\sqrt{1-\big(\frac{h(z_*)}{h(z)}\big(\frac{z}{z_*}\big)^3\big)^2}}\Big]^{1/2}\,. 
	\end{eqnarray}
	We can also write the volume $V(\gamma_A)$ as 
	\begin{eqnarray}
	V(\gamma_A) &=&2\pi^3R^9 L^2 \int_{0}^{z_{*}}\frac{h(z)^{3/2}}{z^4} x(z)dz\,.\label{Vbrane}
	\end{eqnarray}
 We can solve Eq. (\ref{xD3}) exactly, and  obtain
	\begin{eqnarray}
		x(z)=
		\left\{ \begin{array}{lr}
			C_1+\int \frac {h(z_*){z}^{3}}{\sqrt {-{h(z_*)}^{2}{z}^{6}+{z_0}^{6}}}dz \ , & z\leq z_0 \\C_2+\int 
			\frac {h(z_*){z}^{3}{z_0}^{2}}{\sqrt {-{z}^{10}{h(z_*)}^{2}+{z_0}^{10}}} dz\ , &z\geq z_0
		\end{array} \right. \ ,
	\end{eqnarray}
where, $C_{1}$ and $C_{2}$ are integration constants.  The maximal volume, 
which is related to the fidelity susceptibility, is given by 
	\begin{eqnarray}
		&& {V}(\Sigma_{max})=2\pi^3R^9 L^3 \int_{0}^{z_{\infty}}\frac{h(z)^{3/2}}{z^4}\,. dz\label{fidelityD3}
	\end{eqnarray}
Now we will use $h(z)$, and   split the integral into two parts: 
$\int_0^{z_{\infty}}=\int_0^{z_0}+\int_{z_0}^{z_{\infty}}$,  to obtain
	\begin{eqnarray}
		&& {V}(\Sigma_{max})=\frac{-\pi^3R^9 L^3}{3}\,{\frac {{z_{{0}}}^{6}+{z_{\infty}}^{6}}{{z_{\infty}}^{6}{z_{{0}}}^{3}}}\,.\label{VVVVV}
	\end{eqnarray} 
	It may be noted that by    setting $C_{1}=C_{2}=L$, 
the difference of the volumes (\ref{Vbrane}) and (\ref{VVVVV}), is given by  
	\begin{eqnarray}
		V(\gamma_A)-V (\Sigma_{max})=
		\left\{ \begin{array}{lr}
			\int_{0}^{z_0} \frac {h(z_*){z}^{3}}{\sqrt {-{h(z_*)}^{2}{z}^{6}+{z_0}^{6}}}dz \ , & z\leq z_0 \\ \int _{z_0}^{z_{\infty}}
			\frac {h(z_*){z}^{3}{z_0}^{2}}{\sqrt {-{z}^{10}{h(z_*)}^{2}+{z_0}^{10}}} dz\ , &z\geq z_0
		\end{array} \right. \ .
	\end{eqnarray}
Since $h(z_*)>0$, we can express this   as 
	\begin{eqnarray}
		V(\gamma_A)-V (\Sigma_{max})\geq
		\left\{ \begin{array}{lr}
			\frac {z_0h(z_*)}{4} \ , & z\leq z_0 \\
			\frac {h(z_*)}{4{z_0}^{3}}(z_{\infty}^4-z_0^4) \ , &z\geq z_0
		\end{array} \right. \ .
	\end{eqnarray}
So, for a D3-brane, we obtain a     relation between $V(\gamma_A)$ and $V (\Sigma_{max})$,  
$		V(\gamma_A)-V (\Sigma_{max})\geq 0.$ 
However, as the holographic complexity and  fidelity susceptibility  for a  system 
is obtained using 
$V(\gamma_A)$ and $V (\Sigma_{max})$, we obtain the following bound for a D3-brane, 
	\begin{eqnarray}
		\Delta \mathcal{C}_A\geq \Delta \chi_F\,.
	\end{eqnarray} 
So,  we have demonstrated that  for a D3-brane 
the holographic  complexity  is always greater than or equal to than the fidelity susceptibility. 
This was expected as the fidelity susceptibility is calculated using the maximum
volume in the bulk, and the 
holographic complexity is only calculated for a subsystem. 

It is also possible to demonstrate that  a relation exists between the 
holographic complexity and  the entanglement entropy of  D3-brane.  
To obtain this  relation between the holographic complexity and 
entanglement entropy of a D3-brane, we  note that 
$\Delta S_{A}$ is given by 
\begin{eqnarray}
 \Delta S_{A}=\frac{\pi^3R^8L^2}{2G}\int_{0}^{z_{*}}\Big(\frac{h(z)
\sqrt{x'(z)^2+h(z)^2}-\sqrt{x_{AdS}'(z)^2+1}}{z^3}\Big)dz\,, && 
	\end{eqnarray}
and $\Delta\mathcal{C}_A$ is given by 
	\begin{eqnarray}
		&&\Delta\mathcal{C}_A =\frac{\pi^2R^8 L^2}{4G } \int_{0}^{z_{*}}\Big(\frac{h(z)^{3/2}x(z)-x_{AdS}(z)}{z^4}\Big) dz
	\end{eqnarray}
where
\begin{eqnarray}
		x_{AdS}(z)=
		\left\{ \begin{array}{lr}
			C'_1+\int \frac {{z}^{3}}{\sqrt {-{z}^{6}+{z_0}^{6}}}dz \ , & z\leq z_0 \\C'_2+\int 
			\frac {{z}^{3}{z_0}^{2}}{\sqrt {-{z}^{10}+{z_0}^{10}}} dz\ , &z\geq z_0
		\end{array} \right. \ .
	\end{eqnarray}
because $z\sim0$ is the nearly AdS boundary limit. So, now as     $z_*<z_0$, we obtain,  
\begin{eqnarray}
		 \Delta S_{A}&\approx&\frac{\pi^3R^8L^2}{z_{*}^2 G},\\ 
\Delta\mathcal{C}_A&\approx&-\frac{3}{4}\frac{\pi^2 R^8 L^2}{z_{*}^3 G}(C_1-C_1')
	\end{eqnarray}
Total length of this system can be written as 
\begin{eqnarray}
		&&l\approx\frac{4\sqrt{h(z_*)}z_*}{5}
	\end{eqnarray}
By defining the effective holographic temperature $T_{ent}\sim l^{-1}$, 
we obtain the   relation relation between the holographic complexity 
and the holographic entanglement entropy, 
\begin{eqnarray}
&&{\Delta\mathcal{C}_A}= \frac{c\Delta S_{A}}{T_{ent}R}\label{universal}.
\end{eqnarray}
where $c$ is  given by 
\begin{eqnarray}
&&c=\frac{3}{5}\frac{C_1'-C_1}{\pi}\frac{R}{c_{T}}. 
\end{eqnarray}
Here $c_T$ is the proportionality coefficient
in the definition of the  $T_{ent}$ \cite{Bhattacharya:2012mi, Momeni:2015vka},
and $C_1',C_1$, are integration constants. As the only dependence of $c$ on the geometry 
is from the AdS radius $R$, the value of the coefficient $c$ does not depend on the specific 
deformation of the AdS geometry, and so  it can not depend on the specific configuration of 
the D3-branes.  
It may be noted that this bound can also be used to understand the 
meaning of the holographic complexity for a boundary theory, as 
all the other quantities are defined for boundary theory, and thus 
this relation can be used to understand the behavior of  the holographic complexity 
for the boundary theory. 

Thus, we have obtained a relation between the holographic complexity and 
holographic entanglement entropy for a D3-brane. 
However, as the holographic complexity is also related to the fidelity 
susceptibility, we obtain the following holographic bound 
for a D3-brane 
\begin{eqnarray}
\frac{c\Delta S_{A}}{T_{ent}R}=	\Delta \mathcal{C}_A \geq \Delta \chi_F\,.
	\end{eqnarray}
It may be noted that a bound on the holographic entanglement entropy
for a fixed effective holographic temperature can be translated into a bound 
on the holographic complexity, and this in turn can be related to 
a bound on the fidelity susceptibility. So, we have obtained a relation between 
the holographic complexity, holographic entanglement entropy and 
fidelity susceptibility for a D3-brane. The holographic entanglement entropy is 
directly proportional to the holographic complexity, when the 
effective holographic temperature is fixed. Furthermore,  
the holographic complexity is always 
the holographic  complexity  is always greater than or equal
to than the fidelity susceptibility, so the fidelity susceptibility can also be related 
to the holographic entanglement entropy.

 
As it has been recently proposed that the holographic entanglement entropy 
can be  calculated from the action evaluated at a Wheeler-DeWitt 
patch  \cite{prl, prd}, we shall now calculate the 
holographic complexity from action for this D3-brane configuration.   
It may be noted that it is expected that the holographic complexity from 
action will satisfy the bound  \cite{LIoyd}.
This bound has been tested for different AdS black hole geometries \cite{prl, prd, Cai:2016xho},
and we will test it for a D3-brane configuration. 
Now the holographic complexity form action for this D3-brane configuration 
can be obtained by   evaluated the bulk action on 
the Wheeler-deWitt patch.  The full type IIB action can not be used
for such a calculation as no action is known for the    self-dual five form, which 
exists in the full theory.
So, we will    evaluate  
the probe D3-brane action on the Wheeler-de Witt patch, and   not use 
the full type IIB action. In fact,   this solution will depend  on $Q$,
which exists in the probe solution, 
and not the   domain wall solution. So, this only represents the probe 
D3-brane action on the Wheeler-de Witt patch. Now we will  calculate 
the  contributions   of the probe to the complexity from action.
As this quantity is divergence, we will also subtract the background 
AdS contribution from this quantity. Thus, the regularized holographic complexity from the action,
for this D3-brane contribution, can be written as 
\begin{eqnarray}
\Delta \mathcal{C}_W&=& \frac{R^{10}V_3\Omega_5}{\pi \hbar}\Big({\frac {1}{256}}
\,{\frac {{\gamma}^{6}{Q}^{8} ( {z_{{0}}}^{16}-{
\epsilon}^{16} ) }{{\mu}^{12}{r_{{0}}}^{16}}} -\frac{1}{17}\,{\frac {{
\gamma}^{6}{Q}^{9}{z_{{0}}}^{17}}{{\mu}^{13}{r_{{0}}}^{17}}}\nonumber\\  && +{\frac {
17}{36}}\,{\frac {{\gamma}^{6}{Q}^{10}{z_{{0}}}^{18}}{{\mu}^{14}{r_{{0
}}}^{18}}}
-\frac{1}{4}\,{\frac {{Q}^{2}{z_{{0}}}^{2}( {z_{{\infty }}}^{2}-{z_{{0}}
}^{2} ) }{{r_{{0}}}^{4}}}\nonumber \\ &&-\frac{1}{6}\, \Big( -{\frac {35}{2}}\,{\frac 
{{Q}^{6}{\gamma}^{2}}{{\mu}^{4}{r_{{0}}}^{4}}}+{\mu}^{4}{r_{{0}}}^{4}
 \Big( {\frac {35}{2}}\,{\frac {{Q}^{6}{\gamma}^{2}}{{\mu}^{8}{r_{{0}
}}^{8}}}-\frac{1}{8}\,{\frac {{\gamma}^{4}{Q}^{4}}{{\mu}^{8}{r_{{0}}}^{8}}}
\Big)  \Big)\nonumber \\ && \times  {z_{{0}}}^{2} ( {z_{{\infty }}}^{6}-{z_{{0}}}^{
6}) {r_{{0}}}^{-4}{\gamma}^{-2}
\Big), 
\end{eqnarray}
where $\epsilon$ is an IR cutoff and $z_{\infty}$ is the replacement  for a UV cutoff.
It may be noted that unlike the holographic complexity or fidelity susceptibility, 
this holographic complexity from action does not only depend on the geometry, but details 
of the field content of the theory. Thus, it cannot be related to the holographic complexity, 
or fidelity susceptibility, or even holographic entanglement entropy in a direct way. 
This is because these quantities are purely geometric quantities. The main reason for 
this difference is that  unlike entropy, there is an ambiguity in the definition of the complexity, and thus 
many alternative proposals have been made to define complexity of a boundary theory. 
Thus, we cannot relate the holographic complexity from action to those other purely geometric 
quantities. However, we can calculate a different kind of bound for this holographic 
complexity from action. 
Thus, using  the Poincare coordinate $z$, 
such that $z\equiv \frac{r_0}{r},r_0 = \frac{Q}{v}$,
we obtain 
\begin{eqnarray}\label{dA}
 \frac{dA}{dt}&=&R^{10}V_3\Omega_5\int_{0}^{z_{\infty}}
 \frac{h(z)dz}{z}\Big(\frac{1}{\gamma^2}  \phi\left(\frac{r_0}{z}\right)^4 Z +g_s \chi BE
\Big), \nonumber \\
Z&=& \left(1 -
\sqrt{\left( 1 + \frac{\gamma^2 [(\frac{z^2\partial_z\phi}{r_0})^2)^2-E^2]}{\phi(\frac{r_0}{z})^4} \right)
\left(1 +  \frac{\gamma^2 B^2}{\phi(\frac{r_0}{z})^4}\right)} \, \right).
\end{eqnarray}
It has been demonstrated  that the mass of the BPS soltion for this geometry is 
$M = {4\pi Q^2}/{r_0}$ \cite{Schwarz:2014rxa}.  
So, we can write  $
M =4 \pi v Q,  $ and $ r_0={ Q}/{ v }
$, and obtain   
\begin{eqnarray}
&&\frac{d \Delta \mathcal{C}_W }{dt}\approx \frac{0.92040 M}{\pi \hbar}\leq \frac{2M}{\pi \hbar}.
\end{eqnarray}
where chemical potential $v$ is defined through the the coupling constant 
$M_W = gv$. Here  we applied numerical techniques to obtain this holographic bound.  
So, we have demonstrated that for a configuration of D3-branes, 
the holographic complexity from action also satisfies an interesting  holographic bound. 


In this paper, we analyzed certain holographic bounds for D3-brane configurations. 
We analyzed the regularization of the information theoretical quantities dual 
to such a configuration to obtain such bounds.
It may be noted that there are other interesting brane geometries in string theory.
It would be interesting to calculate the holographic complexity, holographic entanglement entropy, and  fidelity susceptibility
for such branes. It might be possible to analyse such holographic bounds for other branes, and geometries that occur in string theory. 
In fact, the argument used for obtaining the relation between the holographic entanglement entropy and holographic complexity of a D3-brane can be easily generalized 
to other geometries. Thus, it would be interesting to analyse if this bound holds for other branes in string theory. In fact, even in M-theory, there 
exist    M2-branes and 
M5-branes, and such quantities can be calculated for such branes. It may be noted that recently,
the superconformal field theory dual to M2-branes has also been obtained, and it is a bi-fundamental 
Chern-Simons-matter theory called the ABJM theory \cite{abjm, abjm0, abjm2}. A holographic dual to the   ABJM   theory with
un-quenched massive flavors has also been studied \cite{abjm1}. 
It is also  possible to mass-deform the ABJM theory \cite{mass}, and the 
   holographic entanglement entropy for the  mass-deformed ABJM theory has been analysed using the AdS/CFT correspondence \cite{mass1}. 
 The holographic complexity for this theory can be calculated using the same minimum surface, and the fidelity susceptibility for this theory 
 can be calculated using the the maximum volume  which
ends on the time slice at the  boundary. It would be interesting to analyse if such a bound exists for the M2-branes.  
 It would also be interest to perform similar analysis for the ABJ theory.
It may be noted that the  fidelity susceptibility   has been used for 
  analyzing   the quantum phase transitions in condensed matter systems \cite{r6,r7, r8}. So, it is possible to holographically analyse the 
  quantum phase transitions using this proposal. It would also be interesting to analyse the consiquences of this bound on the quantum phase 
  transition in condensed matter systems.

\end{document}